\documentclass[letter,traditabstract]{aa} % for the abstract without structuration % (traditional abstract) 
\usepackage{hyperref}
\usepackage{epsf}
\usepackage{graphicx}
\usepackage{txfonts}
\usepackage{natbib}
\usepackage{longtable}
\def\ms{\hbox{\,m\,s$^{-1}$}}         %m.s -1
\def\m2s2{\hbox{\,m$^{2}$\,s$^{-2}$}} %m2.s -2
\def\Mjup{\hbox{$\mathrm{M}_{\rm Jup}$}}
\def\Rjup{\hbox{$\mathrm{R}_{\rm Jup}$}}

\def \sophie{SOPHIE}
\def \kepler{\emph{Kepler}}

\def\kepler{\emph{Kepler}}

\def \1s{$1\,\sigma$}

\def \t0{T$_0$}

\def\Mjup{\hbox{$\mathrm{M}_{\rm Jup}$}}
\def\Rjup{\hbox{$\mathrm{R}_{\rm Jup}$}}
\def \142{KOI-142}

\newcommand{\angstrom}{\mbox{\normalfont\AA}}
\newcommand{\mearth}{{\hbox{$M_{\oplus}$}}}

\begin{document}

\title{SOPHIE velocimetry of \kepler\ transit candidates \thanks{Based on observations collected  with the NASA \kepler\ satellite and with the {\it SOPHIE}  spectrograph on the 1.93-m telescope at Observatoire de Haute-Provence (CNRS), France.}}
            
\subtitle{X KOI-142\,c: first radial velocity confirmation of a non-transiting exoplanet discovered by transit timing}

\author{
S.~C.~C.~Barros \inst{1}
\and R.~F. D\'iaz \inst{1,7}
\and A. Santerne \inst{1,2} 
\and G. Bruno \inst{1}
\and M. Deleuil \inst{1}
\and J.-M. Almenara \inst{1}
\and A.~S. Bonomo \inst{5}
\and F. Bouchy \inst{1}
\and C. Damiani \inst{1}
\and G. H\'ebrard \inst{3,4} 
\and G. Montagnier\inst{3,4}
\and C. Moutou \inst{6,1}}

\institute{Aix Marseille Universit\'e, CNRS, LAM (Laboratoire d'Astrophysique de Marseille) UMR 7326, 13388, Marseille, France
\and Centro de Astrof\'isica, Universidade do Porto, Rua das Estrelas, 4150-762 Porto, Portugal
\and Institut d'Astrophysique de Paris, UMR7095 CNRS, Universit\'e Pierre \& Marie Curie, 98bis boulevard Arago, 75014 Paris, France
\and Observatoire de Haute-Provence, Universit\'e d'Aix-Marseille \& CNRS, 04870 Saint Michel l'Observatoire, France
\and INAF - Osservatorio Astronomico di Torino, via Osservatorio 20, 10025, Pino Torinese, Italy
\and CNRS, Canada-France-Hawaii Telescope Corporation, 65-1238 Mamalahoa Hwy., Kamuela, HI 96743, USA
\and Observatoire Astronomique de l'Universite de Gen\`eve, 51 chemin des Maillettes, 1290 Versoix, Switzerland.}

\abstract{The exoplanet KOI-142b (Kepler-88b) shows transit timing variations (TTVs) with a semi-amplitude of $\sim 12\,$ hours, earning the nickname of ``king of transit variations''. Only the transit of the planet b was detected in the \kepler\ data with an orbital period of $\sim 10.92\,$ days and a radius of $\sim 0.36$ \Rjup.
The TTVs together with the transit duration variations (TDVs) of KOI-142b were analysed by \citet{Nesvorny2013} who found a unique solution for a companion perturbing planet. The authors predicted an outer non-transiting companion, KOI-142c, with a mass of $0.626\pm 0.03$ \Mjup\ and a period of $22.3397^{+0.0021}_{-0.0018}\,$days, and hence close to the 2:1 mean-motion resonance with the inner transiting planet.
We report independent confirmation of KOI-142c using radial velocity observations with the SOPHIE spectrograph at the Observatoire de Haute-Provence. We derive an orbital period of $22.10 \pm 0.25\,$days and a minimum planetary mass of $0.76^{+0.32}_{0.16}\,$\Mjup, both in good agreement with the predictions by previous transit timing analysis. Therefore, this is the first radial velocity confirmation of a non-transiting planet discovered with transit timing variations, providing an independent validation of the TTVs technique.}


\date{Received TBC; accepted TBC}
      
\authorrunning{Barros, S.C.C. et al.}
\titlerunning{RV confirmation of the exoplanet KOI-142\,c:}

\offprints{\\
 \email{susana.barros@lam.fr}}

\maketitle

%
%________________________________________________________________

\section{Introduction}

\citet{Escude2002,Holman2005,Agol2005} have proposed transit timing variation (TTVs) as an additional exoplanet discovery tool. These authors showed that the mutual gravitational interaction of planets close to mean-motion resonances could be strong enough to have a measurable effect on otherwise strictly periodic transit times and would be sensitive to masses lower than the current radial velocity (RV) sensitivity limit.
Despite several ground efforts to detect TTVs on hot-Jupiter systems  \cite[e.g.][]{MillerRicci2008, Gibson2009, Barros2013} their confirmation was only possible with the space-based \kepler\ transiting survey. The first TTV exoplanet system discovered, Kepler-9 \citep{Holman2010}, is composed of pair of transiting Saturn-mass planets near the 2:1 resonance and an inner earth-sized companion. Since then, dynamic analysis of TTVs in \kepler\ transiting multi-planetary systems have allowed a better characterisation of the system and/or helped confirm the planetary nature of many candidates  \cite[e.g.][]{Holman2010,Lissauer2011a, Steffen2012b}.
 However, for cases where only one of the planet transits, it has been shown that, in general, the TTV inversion problem is degenerate \citep{Nesvorny2008,Meschiari2010, Veras2011,Boue2012, Ballard2011}. Only for two special cases, KOI-872 and \142, a unique solution for the companion was found \citep{Nesvorny2012,Nesvorny2013}.

KOI-142b (Kepler-88b) shows large TTVs with a semi-amplitude of $\sim 12\,$hours and a period of $630\,$days \citep{Ford2011,Steffen2012c,Mazeh2013}. TDVs with smaller amplitude (5 minutes) and in phase with the TTVs were also found \citep{Nesvorny2013}. Dynamic analysis of this system led to the prediction of a non-transiting companion, planet c, with a mass of $0.626\pm 0.03$ \Mjup, orbital period of $22.3397^{+0.0021}_{-0.0018}\,$days, inclination of $86.2 \pm 1.5^\circ$ and  eccentricity of $0.05628 \pm 0.0021$  \citep{Nesvorny2013}. Therefore, the second planet is just wide of the 2:1 resonance with the transiting planet. In this configuration, the resonant angle, which measures the displacement of the longitude of the conjunction from the periapsis of the outer planet, circulate (or librate) with a period of 630 days. The high amplitude TTVs and TDVs reflect the resonant angle libration.

 As part of our campaign to validate and characterise \kepler\ transiting exoplanet candidates \cite[e.g.][]{Bouchy2011,Santerne2012} we observed \142\ with the SOPHIE spectrograph at the Observatoire de Haute-Provence. Although we do not detect the known transiting planet in the system, we detect a clear RV signature of a planet companion with a period of $22.10 \pm 0.25\,$days and a minimum mass of $0.76^{+0.32}_{0.16}\,$\Mjup\ in agreement with the predictions of \citet{Nesvorny2013}. In this paper, we present the radial velocity confirmation of KOI-142\,c, making it first radial velocity confirmation of a planet discovered by TTV analysis. We begin by describing the radial velocity observations in Section~2. In Section~3, we present the host star stellar parameters and in Section~4 we give details on our radial velocity fitting procedure and present the results. We finalise with a discussion in Section~5.

\section{\sophie\ spectroscopy and velocimetry}

We obtained 11 spectroscopic observations of \142\ from 1st July 2013 to the 14th November 2013 with the SOPHIE spectrograph mounted on the 1.93m telescope at the Observatoire de Haute-Provence \citep{Perruchot2008, Bouchy2009}. SOPHIE is a thermally-stable high-resolution echelle optical spectrograph fed by a fiber link from the Cassegrain focus of the telescope. The fiber has a diameter of 3" on sky. The observations were obtained in the high-efficiency (HE) mode, that has a resolution R$\sim40\, 000$ and covers a wavelength range of 390-687 nm.

The radial velocities were derived with the SOPHIE pipeline by computing the weighted cross-correlation function (CCF) \citep{Baranne1996, Pepe2002} with a G2 mask. The extracted RVs included corrections due to charge transfer inefficiency of the SOPHIE CCD \citep{Bouchy2009} using the procedure described in \citet{Santerne2012}. SOPHIE HE mode exhibits instrumental variations at long time scales with an amplitude of a few \ms. These were corrected using observations of a bright and stable star, HD\,185144,  obtained on the same nights and with the same instrument setup. Three spectra of \142\ that were affected by the Moon-scattered light were corrected following a procedure similar to the one described by \citet{Bonomo2010}. 

The average signal-to-noise of each spectra is 23 per pixel at 5800 \AA\ and the average RV uncertainty is 11\ms. The radial velocity measurements and the respective uncertainties are available in an online Table. In the same table we also list the exposure time, signal-to-noise per pixel and the bisector span of the cross correlation function. 
The fact that the bisector spans of \142 do not vary significantly, implies that the stellar CCF is not strongly blended with another star. We describe the RV analysis in the Section~\ref{modeling} following the derivation of the host star stellar parameters.

\section{Host star \label{sec.stellarparams}}

After correcting for RV variations and background contamination, all spectra of \142\ were co-added. The signal-to-noise ratio of the resulting spectrum is $\sim 180$ per resolution element at 5800 \AA. The atmospheric parameters were derived using the Versatile Wavelength Analysis package \cite[VWA][]{Bruntt2010a,Bruntt2010b} by minimising the correlation between the FeI and FeII abundances and the excitation potential and equivalent width of the spectral lines. The surface gravity ($\log g$) was measured by imposing that the abundance of FeI and FeII are the same, while also checking consistency with the pressure-sensitive lines {Ca \sc i}  at 6122\angstrom\ and 6162\angstrom. This yielded $T_\mathrm{eff} = 5460 \pm 70$ K, $\log g = 4.6 \pm 0.2$ dex, $[\mathrm{Fe/H}] = 0.25 \pm 0.09$ dex and velocity of microturbulence $v_\mathrm{micro} = 1.3 \pm \, 0.1 \mathrm{km \, s}^{-1}$, implying a G6~V spectral type.  Systematic uncertainties were added in quadrature to the VWA errors estimates for  $T_\mathrm{eff}$ and  $\log g$ following \citep{Bruntt2010b} .
Our measurements are in agreement with previous reported values \citep{Nesvorny2013}. The $v \sin i$ was estimated in conjunction with the macroturbulence velocity $v_\mathrm{macro}$ on a set of isolated spectral lines, yielding $v\sin i = 2 \pm 1 \, \mathrm{km \, s}^{-1}$ and $v_\mathrm{macro} = 1 \pm 1 \, \mathrm{km \, s}^{-1}$. 
The good quality of the combined spectrum of \142\ allowed us to isolate 321 non broadened spectral lines which were used for the analysis. These also permitted us to derive a detailed chemical composition, which is given in an onlin Table.

\section{Radial velocity analysis}
\label{modeling}

\begin{figure*}
\centering
\includegraphics[width=0.45\textwidth]{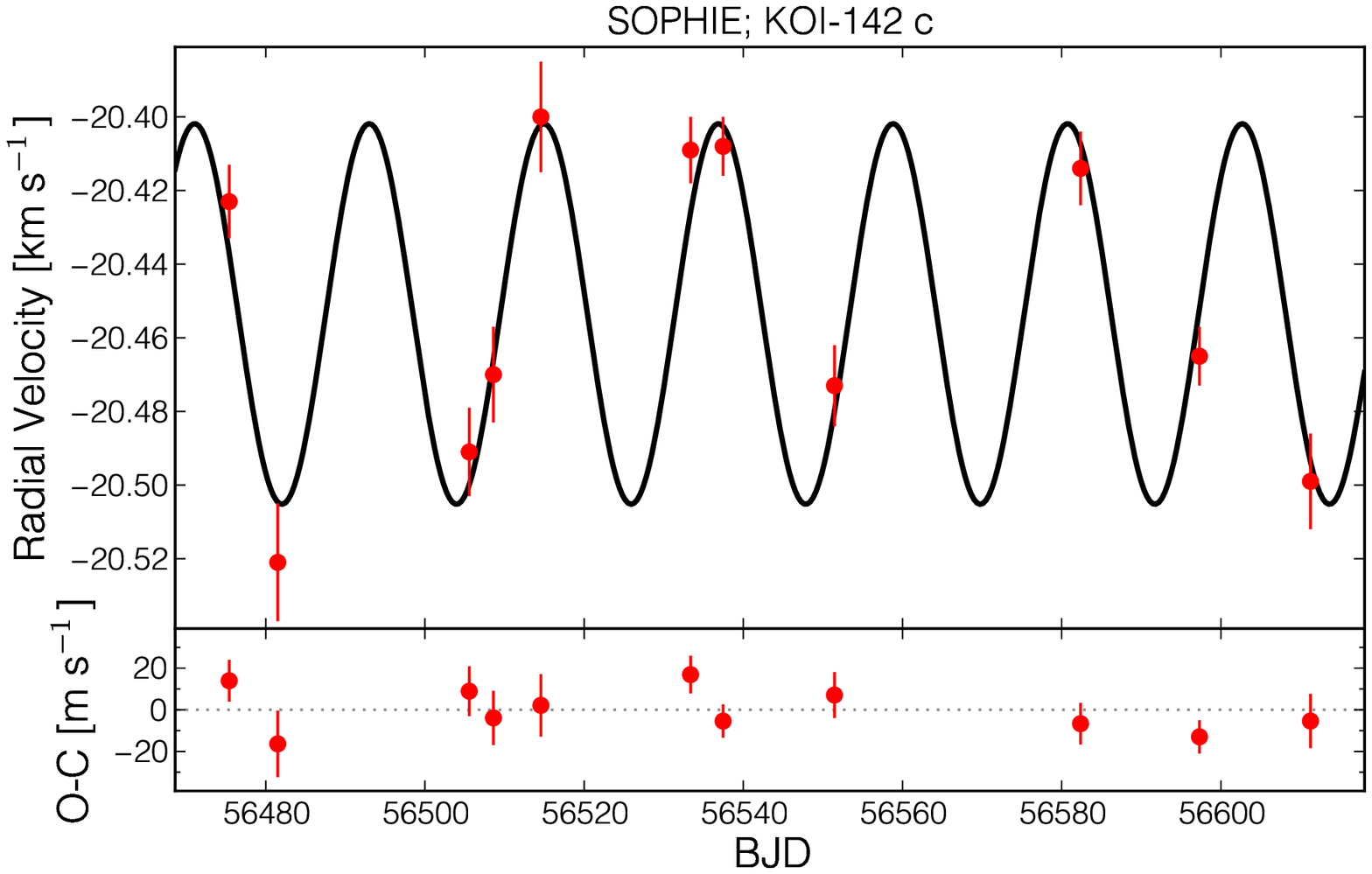}
\includegraphics[width=0.45\textwidth]{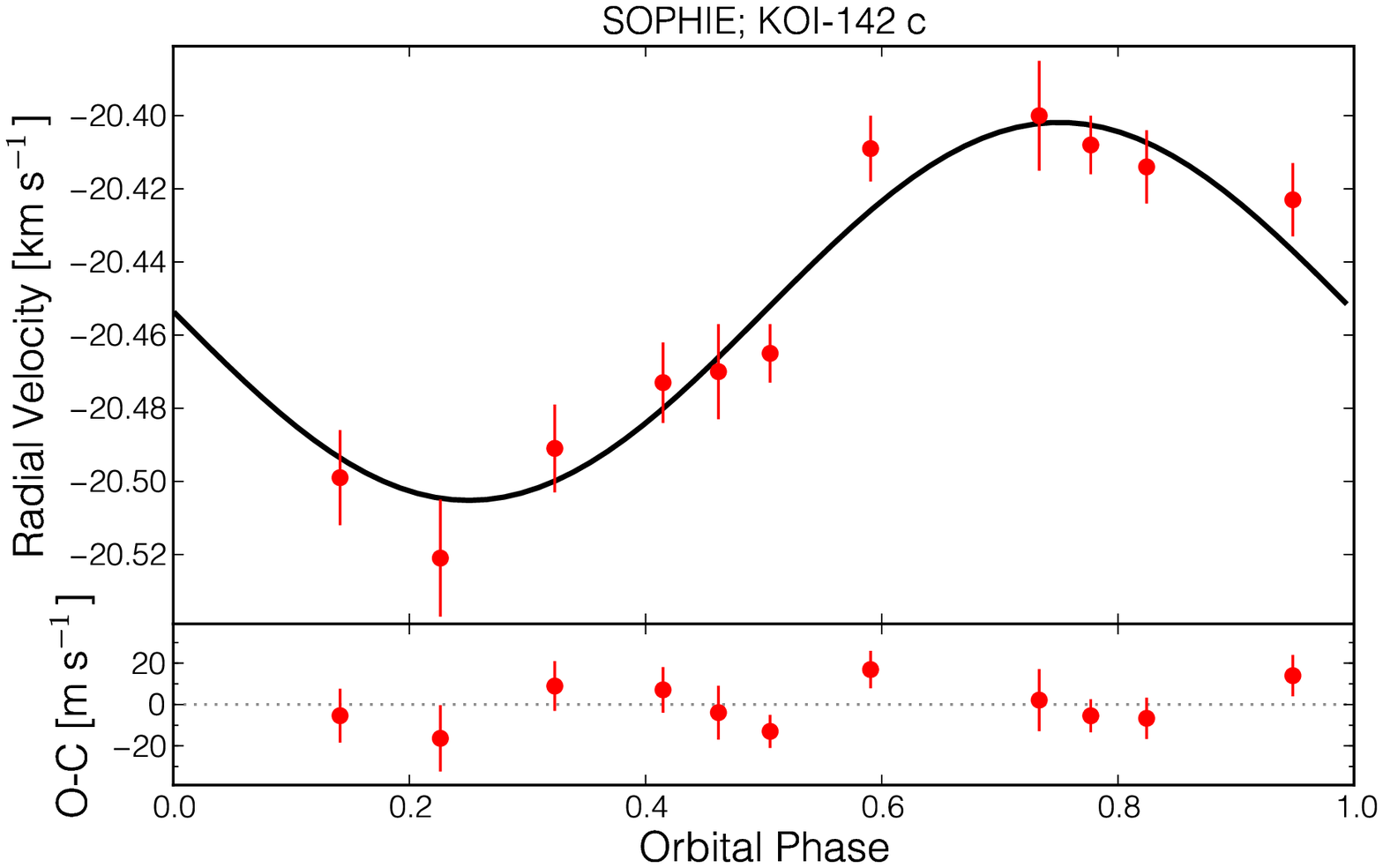}
\caption{SOPHIE radial velocities as a function of time (left) and orbital phase (right), and corresponding residuals. The over-plotted black curve is the most probable fit model. \label{fig.RVdata}}
\end{figure*}

\begin{figure}[t!]
\centering
\includegraphics[width=0.90\columnwidth]{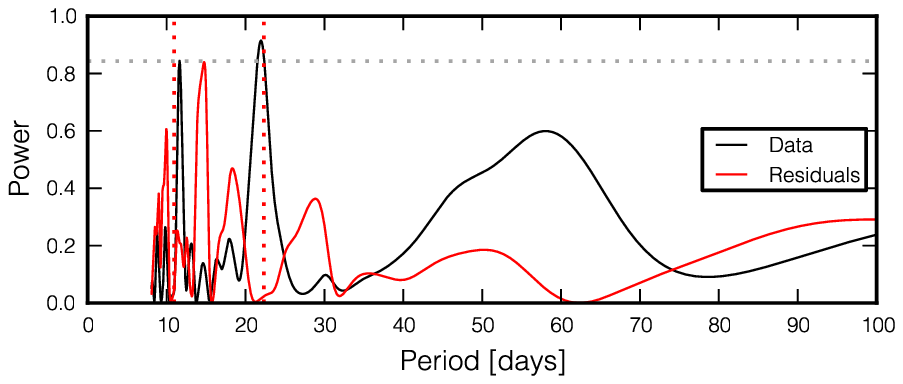}
\includegraphics[width=0.90\columnwidth]{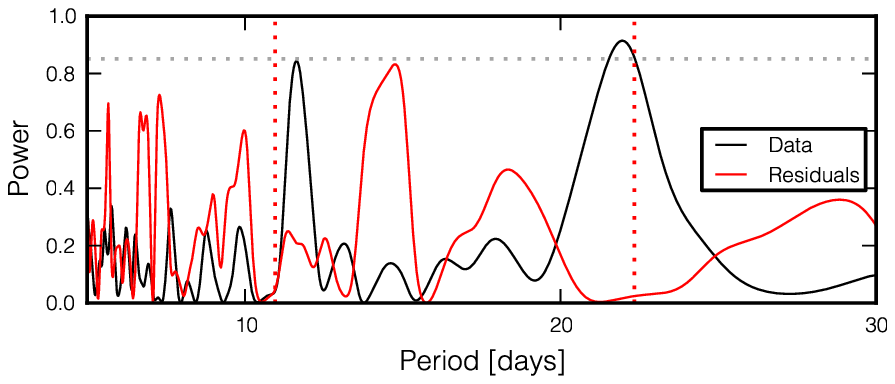}
\caption{Top: Generalised Lomb Scargle periodogram of the SOPHIE radial velocity data. A peak is present at around 22 days. The FAP 10\% level is shown as a dashed horizontal line. The periods reported by \citet{Nesvorny2013} are marked with vertical dashed lines. The Lomb Scargle of the residuals is shown in red. Bottom: Zoom for periods between 5 and 30 days.}
\label{fig.periodogram}
\end{figure}

The radial velocities of \142\ show an amplitude of variation of $\sim 100\,$m/s (Figure~\ref{fig.RVdata}).
A generalised Lomb Scargle periodogram \citep{Zechmeister2009} of the radial velocity measurements show a peak near 22 days, with a false alarm probability (FAP) below 10\% (Figure~\ref{fig.periodogram}).
 This period coincides with the period of the planet discovered by \citet{Nesvorny2013} by transit variation analysis of the KOI-142b planet. The periodogram also shows a lower peak at the first harmonic, with a larger FAP. The Kepler light curve shows a periodicity associated with spot variability at $\sim30\,$days which is not prominent in our RV data.

To check if the 22-day period corresponds to the planet predicted by \citet{Nesvorny2013}, we analysed the SOPHIE RVs of the KOI-142 system using the MCMC algorithm described in detail in \citet{Diaz2013b}.  The RV amplitude expected for the inner planet is $2.58 \pm 0.74$ m~s$^{-1}$ \citep{Nesvorny2013} which would not be detectable with our current data. Therefore, we decided to use a single Keplerian model at the period of the outer, more massive companion. Note that the signal is not expected to be strictly periodic, due to the influence of the inner planet, which changes the orbital parameters of the companion.
The change in the orbital parameters of the putative outer planet has been studied in detail by \citet{Nesvorny2013}, who show that the eccentricity remains well constrained for at least 150 years. A numerical integration of the system up to the epoch of the SOPHIE observations was performed using the hybrid symplectic algorithm implemented in the Mercury6 code \citep{Chambers1999}, and taking as initial conditions the values given in Table 2 of \citet{Nesvorny2013}. We found that over the time of the SOPHIE observations, the orbital elements of KOI-142b change significantly, while the expected change for the proposed companion at 22 days is much smaller\footnote{$\Delta e/e = 0.01$; $\Delta a/a = 3.8\times10^{-4}$; $\Delta i \sim 0.0025^\circ$; $\Delta \omega \sim 0.5^\circ$} than the precision obtained in these parameters (see Table~\ref{table.Params}). Therefore, if not exactly correct, the use of a Keplerian curve to model the signal induced by the second companion is justified over the period of the SOPHIE observations.

Uninformative priors were used in all parameters of the model. In this case, we find a solution at orbital period $P = 22.10 \pm 0.25\,$days, and with eccentricity $e = 0.19^{+0.30}_{-0.14}$. The mode of the posterior eccentricity distribution is at $e \sim 0.065$, with a large tail towards larger values, and a second peak at $e \sim 0.85$. Because of this, most orbital parameters present a similar bimodal distribution. In particular, the posterior of the RV semi-amplitude $K$ also extends up to 277 \ms (95\% confidence level). Using the eccentricity measurement from \citet{Nesvorny2013} as a constraint in our MCMC analysis we measured a RV amplitude of $48.9 \pm 6.0\,$ m~s$^{-1}$, in good agreement with their prediction.

\begin{figure}
\centering
\includegraphics[width=0.75\columnwidth]{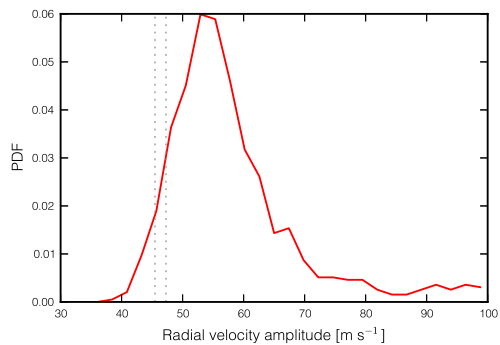}
\caption{Main peak of the marginalised posterior distribution of the RV semi-amplitude. The dashed vertical lines delimit the 1-$\sigma$ range predicted by \citet{Nesvorny2013}. \label{fig.Khist}}
\end{figure}

The inferred mode of the marginal posterior distributions and their 68.3\% confidence intervals are shown in Table~\ref{table.Params}, for the revised stellar parameters described in Section~\ref{sec.stellarparams}. The best-fit model is shown in Figure~\ref{fig.RVdata}. In Figure~\ref{fig.Khist} the marginalised posterior distribution of the radial velocity amplitude is presented. It can be seen that the RV signal is in good agreement with the prediction by \citet{Nesvorny2013}.

\begin{table}[h]
%\vspace{-0.5cm}
\centering
\caption{Parameters for the KOI-142 system at reference epoch ${E = 2456475.40947}$ BJD$_{UTC}$. \label{table.Params}}            
%\hspace{-8cm}
\setlength{\tabcolsep}{3.0mm}
\renewcommand{\footnoterule}{}                          
%\begin{tabular}{p{4.5cm} c}        
\begin{tabular}{l c}        
\multicolumn{2}{l}{\hspace{-0.5cm} Fitted parameters} \\
\hline
Orbital period, $P$ [days]						& $22.10 \pm 0.25$\\
$\sqrt{e}\cos(\omega)$						& $-0.32 \pm 0.25$\\
$\sqrt{e}\sin(\omega)$						& $ 0.24^{+0.17}_{-0.29}$\\
\noalign{\smallskip }
Mean anomaly at epoch [deg]					& $298 ^{+21}_{-77}$\\
\noalign{\smallskip }
RV amplitude, $K$ [m s$^{-1}$] 				& $57.4^{+29}_{-7.5}$\\
\noalign{\smallskip }
Systemic velocity [km s$^{-1}$]					& $-20.4547^{+0.0035}_{-0.0085} $\\
\noalign{\smallskip }
\multicolumn{2}{l}{\hspace{-0.5cm} Spectroscopic parameters} \\
\hline
Effective temperature [K]						& $5460 \pm 70$\\
Surface gravity [dex]							& $4.6 \pm 0.2$\\
Metallicity									& $0.25 \pm 0.09$\\
\multicolumn{2}{l}{\hspace{-0.5cm} Derived parameters} \\
\hline
Periapsis passage, $T_p$, BJD$_{UTC}$		& $2456478.7\pm2.5$	\\
\noalign{\smallskip }
cov($P$,  $T_p$) \tablefootmark{$\dagger$} [days$^2$]					& -0.00323\\	
Orbital eccentricity 							& $0.19 ^{+0.30}_{-0.14}$\\
Semi-major axis (AU) & $0.1529 \pm 0.0021$ \\ 
\noalign{\smallskip }
%Argument of periapsis, $\omega$ [deg]			& $134^{+20}_{-49}$		\\
%\noalign{\smallskip }
Minimum planet mass [M$_\mathrm{J}$] 			& $0.76^{+0.32}_{0.16}$\\
\noalign{\smallskip }
$M_p \sin i / M_s \times 10^4$					& $7.5 ^{+3.3}_{-1.6}$\\
\noalign{\smallskip }
Stellar mass, $M_s$ [M$_\odot$] 				& $0.974 \pm 0.038$\\
Stellar radius, $R_s$ [R$_\odot$] 				& $0.910 \pm 0.040$\\
Age [Gyr] 									& $3.0^{+3.2}_{-2.0}$\\
\noalign{\smallskip}
\hline
\end{tabular}
\tablefoot{
\tablefoottext{$\dagger$}{Covariance between orbital period and time of periapsis passage.} }
\end{table}

A periodogram of the fit residuals does not show any power at the period of the transiting object KOI-142 b. However, it does show a peak at 14.8 days, with a FAP slightly below 10\%. Due to dynamic evolution of the system further RVs are necessary to understand the nature of this peak. One possibility is that it is an harmonic of the stellar rotational period. The residuals of the fit, which are consistent with white noise,  allow to put an upper limit to the mass of the inner planet of $ 53.9 \,$ \mearth\ to 99\% confidence level, which also agrees with the value estimated with TTVs $m_{b}= 8.7 \pm 2.5 \,$ \mearth\  \citep{Nesvorny2013}.

To compute the mass and radius of the stellar host we drew 10,000 samples from the distribution of $T_\mathrm{eff}$ and  $[\mathrm{Fe/H}]$ determined in Section~\ref{sec.stellarparams}, and assumed to follow an uncorrelated multi-normal distribution. We combined them with the stellar density from the transit fit \citep{Nesvorny2013}, corrected for the eccentricity using the posterior samples obtained with the MCMC algorithm. The mass and radius of the star were obtained by interpolation of the Dartmouth stellar tracks \citep{Dotter2008}.
Around 70\% of the samples fall in an unphysical region of parameter space, where the interpolation cannot be done. This is probably due to correlations between $T_\mathrm{eff}$ , $[\mathrm{Fe/H}]$, and the stellar density which were not taken into account in our procedure. This issue  was also encontered by \citet{Nesvorny2013}. With the obtained samples of $M_*$, we computed our posterior distribution for the mass of the planet. These values are reported in Table~\ref{table.Params}.

\section{Discussion}

We present radial velocity observations of \142\ with the SOPHIE spectrograph at the Observatoire de Haute-Provence.
  These measurements allowed us to confirm the discovery of a non-transiting planet in the \142\ system, KOI-142c, in agreement with the TTV predictions of \citet{Nesvorny2013} and providing an independent validation of the TTV method.  The estimated RV semi-amplitude of KOI-142b, $K_b=2.6\,$ m~s$^{-1}$  \citep{Nesvorny2013}, is below the current sensitivity of our RV measurements, and hence the known transiting planet is not included in our analysis. 
We preformed an MCMC analysis to the RVs using a Keplerian model with non-informative priors in all the parameters.
We estimated the orbital period of the c planet, $P_c=22.10 \pm 0.25\,$days and the RV semi-amplitude, $K_c=57.4^{+29}_{-7.5}$ m~s$^{-1}$.
 Therefore, using radial velocity observations we confirm the parameters of KOI-142c predicted by dynamic analysis of transit variations  by \citet{Nesvorny2013}. According to the dynamic TTV analysis, KOI-142c is close to edge-on ($\iota=3.8\pm1.6^\circ$) and hence the planet true mass is expected to be close to the our estimated minimum mass of $0.76^{+0.32}_{0.16}\,$\Mjup.

The confirmation of a TTV prediction is important not just to validate the method but also because unknown (non-transiting) companions could affect the observed TTVs and add further uncertainty to the already complex TTV inversion problem. Another \kepler\ planetary system, KOI-94 \citep{Borucki2011} is one of the few TTV systems where it was possible to constrain  the mass of 3 transiting planets, KOI-94c, KOI-94d and KOI-94e both by RV observations and later by TTVs (no TDVs were measured in KOI-94). However, for the more massive planet, KOI-94d, a large discrepancy between the mass derived from RV measurements and the mass derived from TTVs was found. The RV derived mass, $m_{d}=106 \pm 11$ \mearth\ \citep{Weiss2013}, is approximately the double of the mass predicted from TTVs, $m_{d}=52^{+0.9}_{-7.1}$ \mearth\ \citep{Masuda2013}. These authors suggested the presence of other planets in the system as a possible cause of the discrepancy. The good agreement found for \142\ between the TTV prediction and the RV observations presented here, implies that the TTVs of KOI-142b are not affected by any other possible planets in the system.  Combined RV and TTV analysis of this system will allow to set limits on the presence of other companions and better characterise the planets. These are out of the scope of this paper.

The configuration of multi-planetary systems gives insight into their formation and dynamical evolution \citep{kley&nelson2012}. As many \kepler\ candidates \citep{Lissauer2011b,Fabrycky2012}, \142\ planet pair is just wide of the 2:1 resonance. As mentioned by \citet{Nesvorny2013} the eccentricity of KOI-142c derived from dynamic analysis of TTVs and TDVs, $e_c=0.05628 \pm 0.0021$, is larger than expected by gravitational interaction of the two planets \citep{Lee2013}. Moreover, tidal damping can only explain this system if the dissipation efficiency of the inner planet was $\sim 6.5$ higher than expected for a planet of this type \citep{Lee2013}. However, since planet c is capable of opening a gap on the disc, planet-disc interaction \citep{Baruteau2013}, or planet-planet interaction within the disc \citep{Lega2013} might be able to explain the current configuration. Further RV observations together with dynamic analysis of the system will bring valuable insight into the formation history of this remarkable system.

\begin{acknowledgements}
We thank the staff at Haute-Provence Observatory. We acknowledge the PNP of CNRS/INSU, and the French ANR for their support. The team at LAM acknowledges support by grants  98761 (SCCB), 251091 (JMA) and 426808 (CD). RFD was supported by CNES via the its postdoctoral fellowship program.
AS acknowledges the support of the European Research Council/European Community under the FP7 through
Starting Grant agreement number 239953. ASB gratefully acknowledges support through INAF/HARPS-N fellowship.
\end{acknowledgements}

\bibliographystyle{bibtex/aa} 
\bibliography{susana}

\end{document}